\pdfoutput=1

\documentclass{mwave}

\usepackage{amsmath}

\shorttitle{Rotational Components}
\shortauthors{Sheeley}

\graphicspath{{./}}

\begin{document}

\title{Rotational Components of the Sun's Mean Field}

\author[0000-0002-6612-3498]{Neil R. Sheeley, Jr.}
\affiliation{Visiting Research Scientist\\
Lunar and Planetary Laboratory, University of Arizona \\
Tucson, AZ 85721, USA}

\begin{abstract}
This paper uses wavelet transforms to look for the rotational frequencies
of the Sun's mean line-of-sight magnetic field.  For a sufficiently high wavelet frequency, the
spectra of the dipole, quadrupole, and hexapole field components each show a
time-dependent fine structure with periods in the range of 26.5-30 days and their harmonics.
These maps confirm that a large enhancement of 30-day power occurred in the dipole field
during 1989-1990, as recorded previously using Fourier techniques  \citep{Sheeley_2022}.  Also,
during some years the maps show power at 26.5 days (or its harmonics), that is clearly
distinguishable from the 26.9-27.0 day rotation period at the Sun's equator.  In at least one case,
the 26.5-day period was a wave phenomenon caused by the systematic eruption of active regions
at progressively more western locations in the Carrington coordinate system, as if the flux were
emerging from a fixed longitude in a faster rotating subsurface layer.  Based on previous studies
of the mean field \citep{SHEDB_1985,SDV_1986,Sheeley_2022}, I conclude that the enhanced
wavelet patterns in this paper are regions where magnetic flux is emerging in configurations
that strengthen the Sun's horizontal dipole, quadrupole, and hexapole fields, and (in the case of the
more slowly rotating patterns) where this flux is being transported to mid-latitudes whose rotation
periods are in the range 28-30 days.
\end{abstract}


\section{Introduction} \label{sec:intro}

In a previous paper, I used Fourier transforms of Wilcox Solar Observatory (WSO) measurements
from 16 May 1975 to 16 November 2021 to study the rotational components of the Sun's mean
line-of-sight magnetic field \citep{Sheeley_2022}.  With this 46.5-yr sequence of daily
measurements, the transforms resolved long-lived 2-sector recurrence patterns of
27, 28.5, and 30 days, and showed the existence of 4-sector and 6-sector patterns.  For
suitable choices of the limits of integration, it was possible to determine the temporal origin
of some of the 2-sector recurrence patterns.

Now, in this paper, I present the results of applying wavelet transforms to the WSO
mean-field measurements updated though 5 July 2022.  These transforms are designed
to provide spectral power as a function of time and wavelet scale (equivalent to spectral
frequency, ${\omega}$, or period, $T=2{\pi}/{\omega}$).   The
analysis differs from an earlier wavelet analysis of the WSO data by
\cite{BOB_2002}.   They considered the general properties of oscillating structures over a
broad range of periods from 90 minutes to 11 years, whereas in this study, I am focussing on the
detailed spectral properties for periods comparable to the 27-day rotational period of the
Sun and its second and third harmonics at 13.5 days and 9 days.

By selecting the wavelet frequency sufficiently high, I have been able to
resolve the components with periods of $\sim$27 days from those of periods 28.5 days
and 30 days.   In addition, the resulting two-dimensional maps show resolved spectral
power, not only for the horizontal dipole fields with azimuthal number $m=1$ found in
the earlier paper \citep{Sheeley_2022}, but also for the horizontal quadrupole ($m=2$)
and hexapole ($m=3$) fields.  As we will see, this approach verifies my previous results
using Fourier transforms.  Also, it reveals a long-lived quadrupole pattern in 1990-1991,
having a period that is shorter than the equatorial rotation period at the Sun's
surface ($\sim$26.5 days, compared to $\sim$26.9-27.0 days of synodic rotation).

After describing the wavelet technique, I will show the resulting maps of
spectral power (the so-called wavelet scaleograms), and present a possible interpretation of
this 2-dimensional power.   Appendices A and B contain derivations of mathematical
relations used in the text, and Appendices C and D contain derivations of the mean field and
open flux of an idealized magnetic doublet representing the flux in a nominal bipolar
magnetic region.

\section{Analysis Techniques}\label{sec:analysis_techniques}

\subsection{Range of Frequencies}
The WSO mean-field used in this paper consists of a single measurement every day
for $N=17,218$ days, corresponding to the 47-yr interval from 16 May 1975 to
5 July 2022.  The corresponding range of frequencies can be interpreted as follows:
The maximum frequency, ${\omega}_{max}$ (the Nyquist frequency), is determined
by one point every half wave.  Consequently, ${\omega}_{max}={\pi}$ rad point$^{-1}$, which is
${\pi}$ rad day$^{-1}$ at the observing rate of 1 point day$^{-1}$.  The minimum frequency,
${\omega}_{min}$, is determined by putting $N/2$ points in the half wave, so that
${\omega}_{min}={\pi}/(N/2)=2{\pi}/N$ rad day$^{-1}$.

Thus, we can express the range, R, in rad day$^{-1}$ as
\begin{equation}
R~=~\left (\frac{2{\pi}}{N},~{\pi} \right )~=~ \frac{2{\pi}}{N} \left (1, ~\frac{N}{2} \right )~=~
{\omega}_{min}  \left (1, ~\frac{N}{2} \right ).
\end{equation}
The smallest frequency, ${\omega}_{min}=2{\pi}/N$, is also the spectral resolution,
${\Delta}{\omega}$, of the data.  In terms of this resolution, the full set of
frequencies, ${\Omega}$, becomes
\begin{equation}
{\Omega}~=~{\Delta}{\omega} \left \{ 1, ~2, ~3, ~...,~ (\frac{N}{2}-1), ~ \frac{N}{2} \right \}.
\end{equation}

Of course, it was not possible to obtain mean-field measurements on every day due to a variety
of conditions that presumably included adverse weather, and occasional hardware and software
problems.  Consequently, there were data gaps on 3069 of the 17218 days, corresponding to
$17.8\%$ of the data.  On these days, I set the mean field equal to zero before computing
the wavelet transforms.  To get a feeling for how that may have affected the resulting maps,
I also replaced those values by random numbers in the range (-0.32, +0.32) G, where 0.32 G
was the root-mean-square value of the other measurements.  This had essentially no effect
on the distribution of wavelet power, and even when the threshold was set higher than 0.32 G,
only the faint background distribution was affected.  The enhanced regions of power were
unchanged.  I did not try interpolating between adjacent non-gap values because the gaps
were not always isolated days, but came in a variety of combinations ranging from one to
several days and sometimes to a few weeks or more. 

\subsection{The Wavelet and the Wavelet Transform}
When using Fourier techniques, one combines the time series, f(t), with an oscillating
factor, $e^{i{\omega}t}$, that spans the entire time series.  This gives the power as
a function of frequency, ${\omega}$, but does not indicate when the spectral peaks
of this distribution originated.  With wavelets, one introduces a damping factor of
the form, $e^{-(1/2)\{(t-{\tau})/s\}^2}$,  to limit the range of oscillation to a temporal scale, s, around
the time $t={\tau}$, and  then combines the time series, f(t), with this damped and phase-shifted
oscillation to obtain power as a function of the temporal shift, ${\tau}$, and the temporal scale, s.
In particular, the  damped, but unshifted wavelet, ${\psi}$, analogous to the Fourier factor,
$e^{i{\omega}t}$, is
\begin{equation}
{\psi}~{\sim}~ e^{i{\omega}_{0}t}~e^{-\frac{1}{2}(t/s)^2}~{\sim}~
~  e^{i{\gamma}(t/s)}~e^{-\frac{1}{2}(t/s)^2},
\end{equation}
where ${\gamma}$ is a dimensionless constant, indicating the number of radians of oscillation
in a scale length, s.  Equivalently, ${\gamma}/2{\pi}$ is the number of oscillations in a decay time
scale, s.  Now,  ${\omega}_{0}$ is a true frequency in rad day$^{-1}$, given by
\begin{equation}
 {\omega}_{0}~=~\frac{{\gamma}}{s}.
 \end{equation}
This differs from the conventional notation, $e^{i{\omega}_{0}(t/s)}$, in which ${\omega}_{0}$ 
is regarded as a `dimensionless frequency',  and is usually taken to be a number approximately
equal to 6 \citep{BOB_2002,POD_2009,TOR_98,SHI_2022}.  I prefer to
assign another variable, ${\gamma}$, to the dimensionless quantity and to reserve the symbol,
${\omega}_{0}$, for a bonafide frequency with dimensions of rad day$^{-1}$.  For simplicity, I have omitted a normalization factor of ${\pi}^{-1/4}$ from these wavelet expressions, but I will include it
when the actual calculations are performed.  The idea is to shift this wavelet by an amount, ${\tau}$,
to form
\begin{equation}
{\psi} \left (\frac{t-{\tau}}{s} \right )~{\sim}~e^{i{\gamma} ((t-{\tau})/s)}~e^{-\frac{1}{2}((t-{\tau})/s)^2},
\end{equation}
and then find the value of 
\begin{equation}
F(t,s)~=~\frac{1}{\sqrt{s}}\int_{-\infty}^{+\infty}B_{m}(\tau){\psi}^{*}(\frac{t-{\tau}}{s})d{\tau}~=~
\frac{{\pi}^{-1/4}}{\sqrt{s}} \int_{-\infty}^{+\infty}B_{m}({\tau})~e^{-\frac{1}{2}\{(t-{\tau})/s\}^2}
e^{-i{\gamma}\{(t-{\tau})/s\}}d{\tau},
\end{equation}
where $B_{m}({\tau})$ is the mean-field time series, and the factor of ${\pi}^{-1/4}$ has now been included in the wavelet.   Also, I have interchanged the roles of $t$ and ${\tau}$ so that $t$ survives
the integration and now refers to the time shift of the wavelet from its peak value.  Because $F(t,s)$ is usually a complex number,
we plot the real number, $F(t,s)F^{*}(t,s)=|F(t,s)|^2$, as a function of $t$ (on the horizontal axis)
and as a logarithmic function of spectral period, $T=2{\pi}/{\omega}$, (on the vertical axis).
In this case, the vertical axis indicates $\log_{2}(2T)$ increasing downward from 1 at the top of the
map.

As we shall see in the next section, it will be necessary to choose ${\gamma}>>6$ to resolve the individual solar rotation periods and see the effects of differential rotation.  As shown in
Appendix A, for such very large values of ${\gamma}$, there is essentially no difference between
${\omega}_{0}$, the frequency defined by Eq(4), and ${\omega}$, the frequency of the oscillating
mean field, obtained from the wavelet analysis.  This means that there is no need to retain the 0
on ${\omega}_{0}$ or on $T_{0}=2{\pi}/{\omega}_{0}$.   Finally, as a point of terminology,
${\psi}$ is a Gabor wavelet that depends on ${\gamma}$, and for the special case of
${\gamma}={\pi}{\sqrt{2/ln2}}~{\approx}~5.336$, the Gabor wavelet reduces to what is usually
called a Morlet wavelet.
 
\section{Results}  
Figure~1 compares the wavelet power in the WSO mean field (top panel)
with the monthly averaged sunspot number from the Royal Observatory of Belgium (SILSO)
shown in the lower panel.
\begin{figure}[h!]
 \centerline{
 \fbox{\includegraphics[bb=20 150 590 640,clip,width=0.90\textwidth]
 {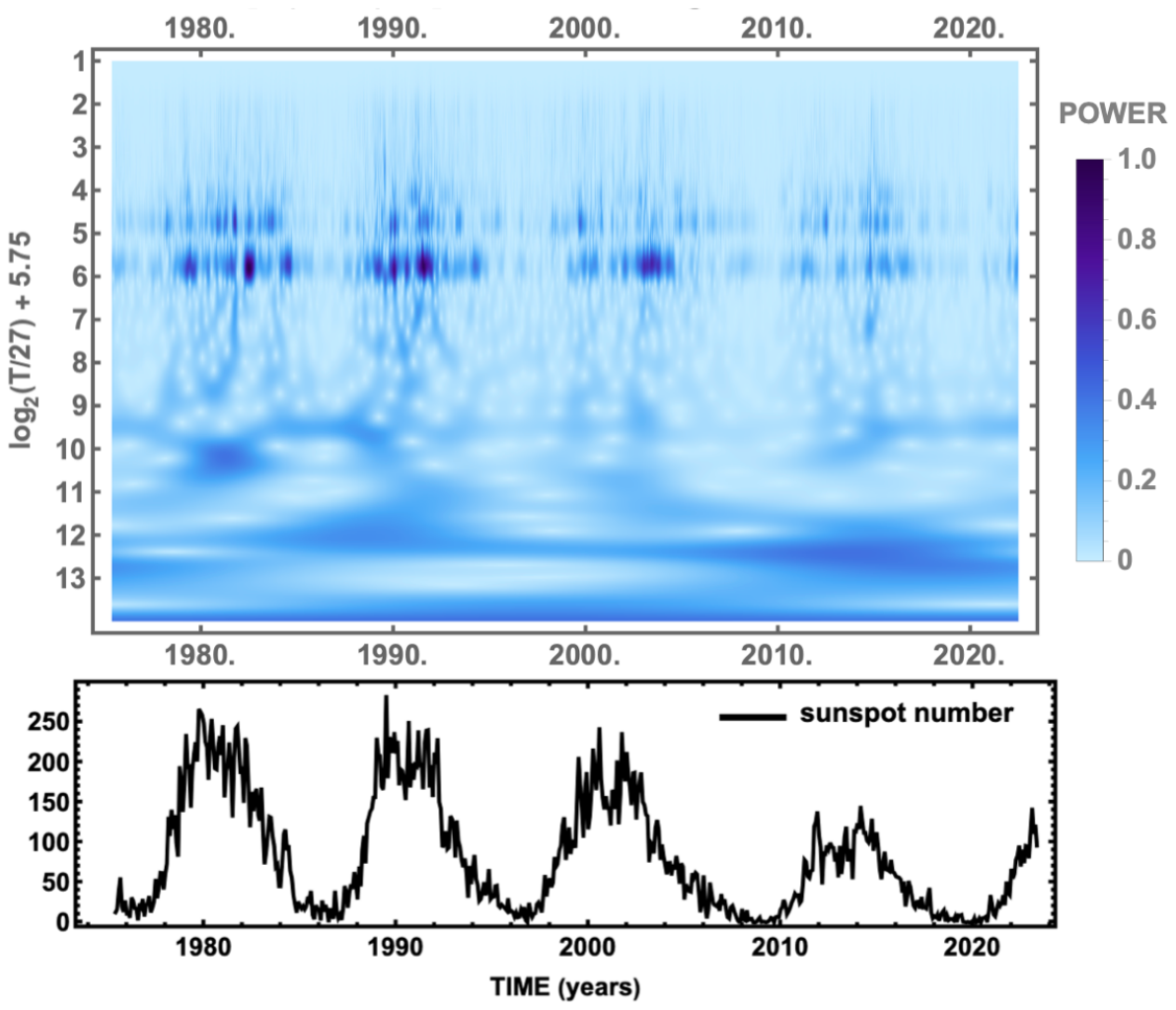}}}
\caption{(top) Wavelet scaleogram of the WSO mean field measurements, showing power
as a function of time, t,
in years, and period, T, in days expressed logarithmically as $\log_{2}(T/27)+5.75$.   Here,
${\gamma}={\pi}{\sqrt{2/ln2}}~{\approx}~5.33$, corresponding to a Morlet wavelet.
(bottom) Monthly averaged sunspot number from the Royal Observatory of Belgium (SILSO).}
\label{fig:fig1}
\end{figure}
Here, ${\gamma}={\pi}{\sqrt{2/ln2}}~{\approx}~5.33$, corresponding to a Morlet wavelet
transform of the WSO mean-field measurements.   This is essentially what \cite{BOB_2002}
obtained for the WSO data through 2001 using ${\gamma}=6$.  (See their Figure 1).

The oscillation period, $T$, is indicated in powers of 2  (like octaves of a musical scale), running logarithmically downward along the vertical axis as $\log_{2}(2T)$.   For
convenience, I use the approximation that $\log_{2}(2T)~{\approx}~{\log}_{2}(T/27) + 5.75$
(equivalent to $\log_{2}27~{\approx}~4.75$), which shows directly that the 27-day equatorial
rotation period occurs at $5.75$ on the vertical axis.  Also, I used 32 steps per octave so that
\begin{equation}
{\log}_{2}(2T)~=~1~+~{\log}_{2}T~=~1~+~\frac{j}{n},
\end{equation}
with $n=32$ and $j$ running from 1 to 416 (13${\times}$32).  In this case, $j=416$ gives
${\log}_{2}(2T)=14$.   Note that ${\log}_{2}N={\log}_{2}17218~{\approx}~14.07$, so that the ordinate values from 1 to 14 almost span the full set of 17218 measurements.

The map shows a variety of spectral and temporal features.   Toward the bottom of the map where ordinate values are in the range 11-13 (corresponding to times of years), coarse structures are aligned horizontally.  Near the top of the map where ordinate values are in the range 4-7 (8-64 days),  fine structures are aligned vertically.  In between, the transition seems to consist of a multitude of
round features scattered through the ordinate range of $\log_{2}2T=7-9$, corresponding to
periods in the range 64-256 days.

 Looking more closely, we can see that the vertical fine structures are
distributed in rows running horizontally at locations of 5.75 (27 days), 4.75 (27/2 days), and
more faintly at 4.16 (27/3 days), corresponding to the equatorial rotation period and its
second and third harmonics.  These features indicate rotational contributions from the dipole,
quadrupole, and hexapole moments of the Sun's field.  They fall into
four time intervals corresponding to sunspot cycles 21-24.  Their intensities
are roughly correlated with the strengths of the sunspot cycles and are relatively weak during
cycle 24.  However, these features are widely distributed over each sunspot cycle, and often occur
early in the declining phase of the cycle, as \cite{Sheeley_2015} found in their analysis of the
Sun's large-scale field.

If we represent the vertical dimension of this map by $y=\log_{2}(2T)$, then we can relate small
changes in $y$ to changes in the period, T.  We do this by converting to natural
logarithms and then differentiating to obtain
\begin{equation}
dy~ln2~=~\frac{dT}{T}.
\end{equation}
Thus, small changes of $y$ are proportional to the fractional change ${\Delta}T/T$.
In Figure~1, the rotational fine structures have a vertical extent of about ${\Delta}y = 0.5$, corresponding to ${\Delta}T/T=0.5~ln2 = 0.346$, which is ${\Delta}T=9.4$ days at $T=27$ days
and 4.7 days at $T=27/2$ days.  So ${\gamma}=5.33$ gives a relatively coarse picture of these
rotational features.  In other words, the Morlet transform shows the dipole, quadrupole and hexapole components with good temporal resolution (${\lesssim}~1$ solar rotation period), but it does not show the rotational fine structures of these components.  On the other hand, with 32 pixels per octave (corresponding to $dy=1/32$ and $dT/T=0.02$), the display is capable of resolving features whose periods differ by $dT=0.54$ days, which includes the 27-day, 28.5-day, and 30-day periods.

However, to achieve this spectral resolution, it is necessary to increase ${\gamma}$, or
${\gamma}/2{\pi}$, which is the number of waves of period, T, in a decay time, s, as indicated
by Eq(4) rewritten in the form
\begin{equation}
s~=~(\frac{{\gamma}}{2{\pi}})T.
\end{equation}
To understand this, recall from section 2.1 that
a Fourier transform of the entire set of $N=17218$ points has a frequency
resolution of $1/N$ cycles day$^{-1}$.  In other words, the frequency resolution is
inversely related to the number of data points in the sample.  Accordingly, we would
expect wavelets of scale $s$ to give a frequency resolution ${\sim}1/s$ cycles day$^{-1}$.
As shown in Eq(B18b) of Appendix B, the root-mean-square angular resolution is
$({\Delta}{\omega})_{rms} = 1/(s{\sqrt2})$.  Dividing by the angular frequency, ${\omega}$,
given by Eq(4), we obtain 
${\Delta}{\omega}/{\omega}=1/({\gamma}{\sqrt2})$. Also, because
${\Delta}{\omega}/{\omega}=-{\Delta}T/T$, it follows that the spectral resolution is
\begin{equation}
|\frac{{\Delta}{\omega}}{{\omega}}|~=~|\frac{{\Delta}T}{T}|~=~\frac{1}{{\sqrt2}}\frac{1}{{\gamma}}.
\end{equation}

For a given period, $T=27$ days, we can obtain a large value of $s$ (corresponding to
a high spectral resolution) by choosing a large value of ${\gamma}$.
As a specific case, let's take ${\gamma}/2{\pi}=20$, corresponding to $|{\Delta}T/T|=0.0056$ and
${\Delta}T=0.15$ days.  Then, using Eq(9) and Eqs(B18a), we obtain
\begin{equation}
({\Delta}t)_{rms}~=~\frac{s}{\sqrt2}~=~\frac{1}{\sqrt2}(\frac{{\gamma}}{2{\pi}})T~=~\frac{1}{\sqrt2}{\times}20{\times}27~\text{days}~=~381.8~ \text{days}~=~1.04~ \text{yr}.
\end{equation}
Thus, for this example of $T=27$ days and ${\gamma}/2{\pi}=20$, the spectral resolution
is 0.15 days, which is sufficient to distinguish the 27-day, 28.5-day, and 30-day spectral
features, but is obtained at the cost of degrading the temporal resolution to about 1 yr.

Figure~2 illustrates this effect by changing ${\gamma}/2{\pi}$ in powers of 2 ranging
from 1 in the top panel to 16 in the bottom panel. 
\begin{figure}[h!]
\begin{interactive}{animation}{slow_then_reverse.mp4}
figure call
 \centerline{
 \fbox{\includegraphics[bb=20 30 630 205,clip,width=0.80\textwidth]
 {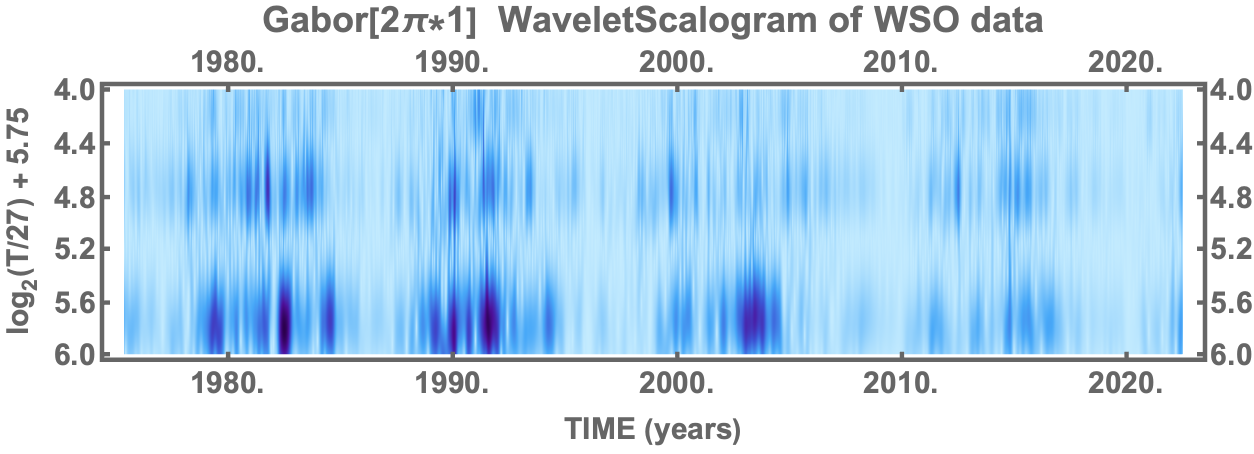}}}
 \centerline{
  \vspace{0.01in}
  \fbox{\includegraphics[bb=20 30 630 188,clip,width=0.80\textwidth]
 {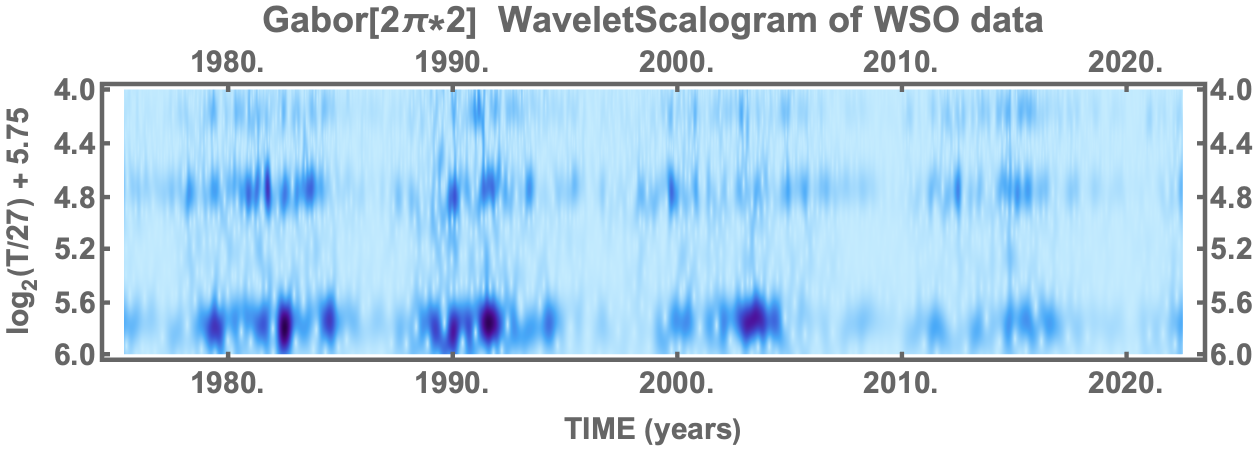}}}
  \centerline{
  \vspace{0.01in}
  \fbox{\includegraphics[bb=20 30 630 188,clip,width=0.80\textwidth]
 {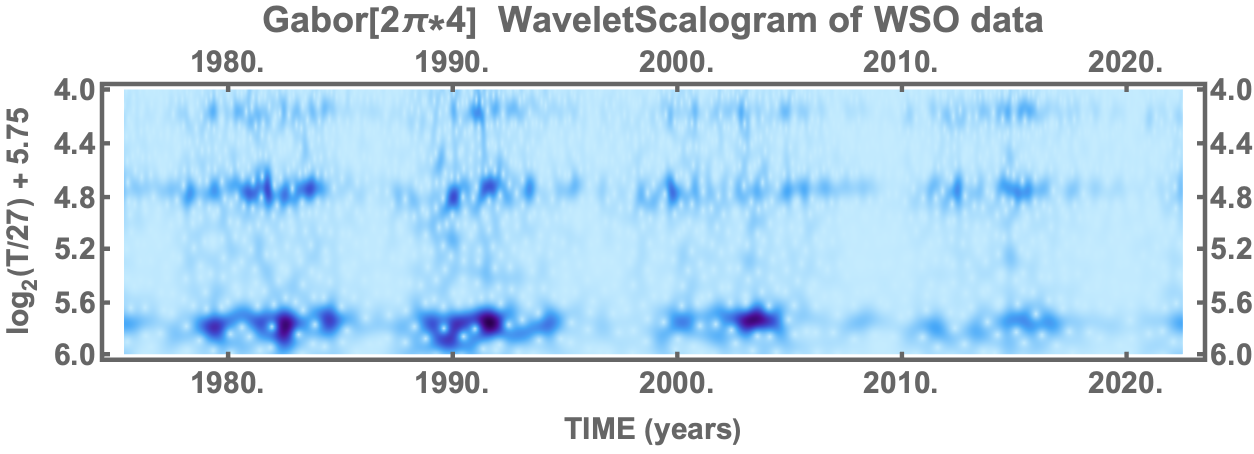}}}
  \centerline{
  \vspace{0.01in}
  \fbox{\includegraphics[bb=20 30 630 188,clip,width=0.80\textwidth]
 {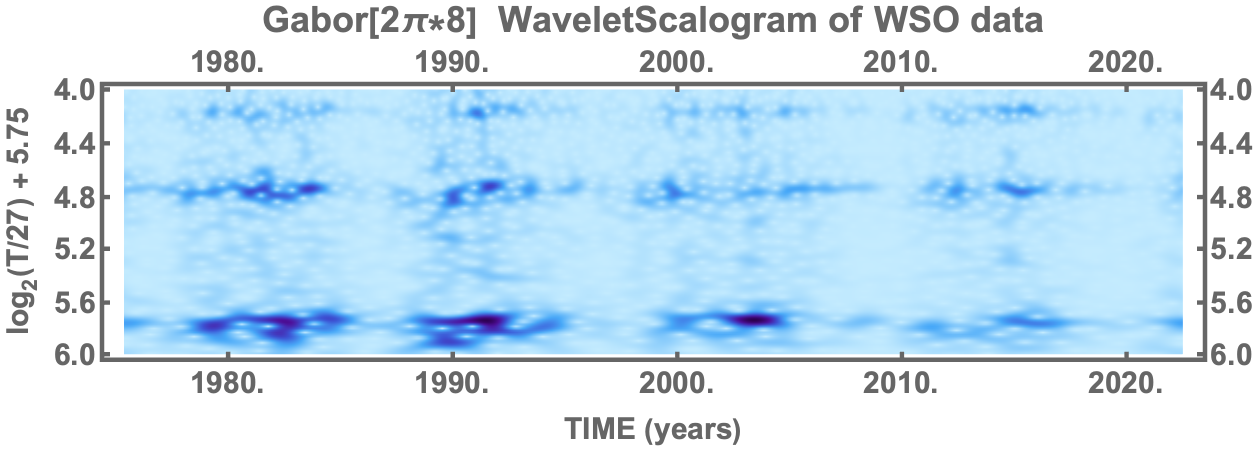}}}
  \centerline{
  \vspace{0.01in}
  \fbox{\includegraphics[bb=20 03 630 188,clip,width=0.80\textwidth]
 {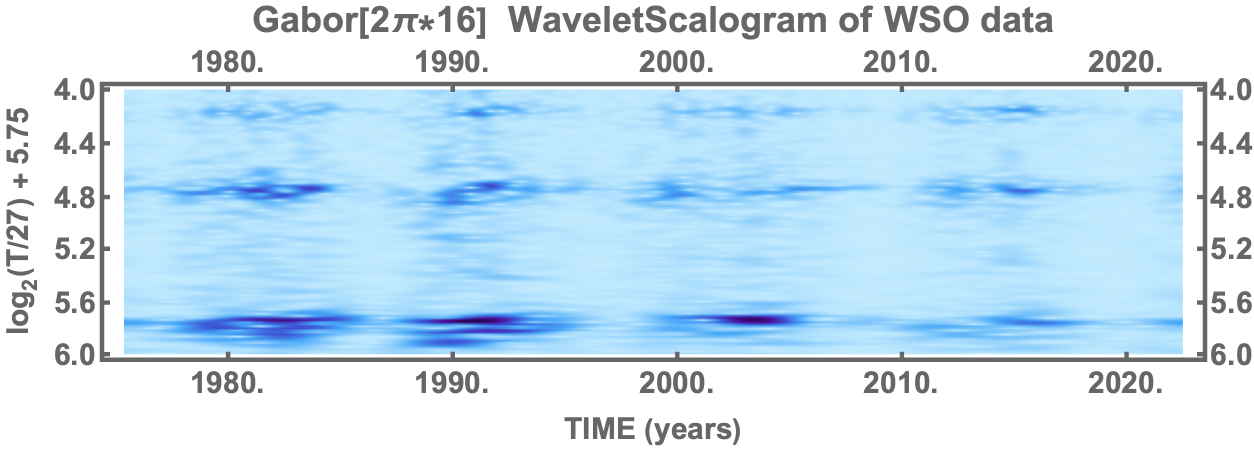}}}
 \end{interactive}
\caption{A sequence of Gabor scaleograms for ${\gamma}/2{\pi}=1$ (top panel) to 16
(bottom panel) in steps of a factor of 2, showing the transition from high temporal
resolution (on the horizontal axis) to high spectral resolution (on the vertical axis).
The vertical axis is a logarithmic indication of spectral period, T, expressed in days as
${\log}_{2}(2T)~{\approx}~{\log}_{2}(T/27)+5.75$. }
\label{fig:fig2}
\end{figure}
\noindent
In the top panel, the narrow slivers of power have widths on the order of a 27-day
solar rotation or less.  But their vertical extensions are about 0.5 unit, corresponding
to several days of frequency resolution, as mentioned above.  In the bottom panel where
${\gamma}/2{\pi}=16$,
the spectral features are aligned in horizontal strips of widths ${\lesssim}$ 0.03 unit
(corresponding to about 0.5 day) and separations of about 0.1 unit (corresponding to
${\approx}~2$ days, so that the rotation periods of 27, 28.5, and 30 days are clearly resolved.
These three components are especially noticeable in 1990 when the dipole field has significant
power at all three frequencies.  

By comparison, in the second panel from the top where ${\gamma}/2{\pi}=2$, the distribution of
wavelet power looks very much like the temporal plots of power in the $m=1$, $m=2$, and $m=3$ modes, as shown in Figure~7 of \cite{Sheeley_2022}.
This is as one would expect because those plots were obtained by inverting the Fourier
transform of the entire data set for the three frequency bands corresponding to $m=1$, $m=2$, and
$m=3$.

I used maps like those in Figure~2 to make a movie of the wavelet power, but with a wider
and finer range of ${\gamma}/2{\pi}$.  When viewed in a single-step mode, the movie allows
one to investigate the transition from high temporal resolution to high spectral resolution in detail.
The idea is to identify individual features and track them backward and forward as a function of
${\gamma}/2{\pi}$.  In this way, one might be able to identify the separate eruptions of flux that
contribute to the long-term patterns of the mean field.  Figure~2 is sufficient to make these
identifications for some simple cases.  For example, the 30-day, 2-sector pattern
in 1989-1990 seems to have originated in two bursts, one around April 1989 and the other in 
January 1990.  Likewise, the 27-day, 2-sector pattern in 2003-2004 seems to have originated
in several eruptions during that time.  
Further discussion of this transition between high temporal resolution and high
spectral resolution is contained in Appendix B.

Figure~3 shows the wavelet power in the dipole (m=1), quadrupole (m=2), and
hexapole (m=3) regions of the spectrum near 5.75, 4.75, and 4.16 on the
vertical axis.  In the middle and upper panels, the values of ${\gamma}/2{\pi}$
have
\begin{figure}[h!]
 \centerline{
 \fbox{\includegraphics[bb=40 85 565 725,clip,angle=90,width=0.96\textwidth]
 {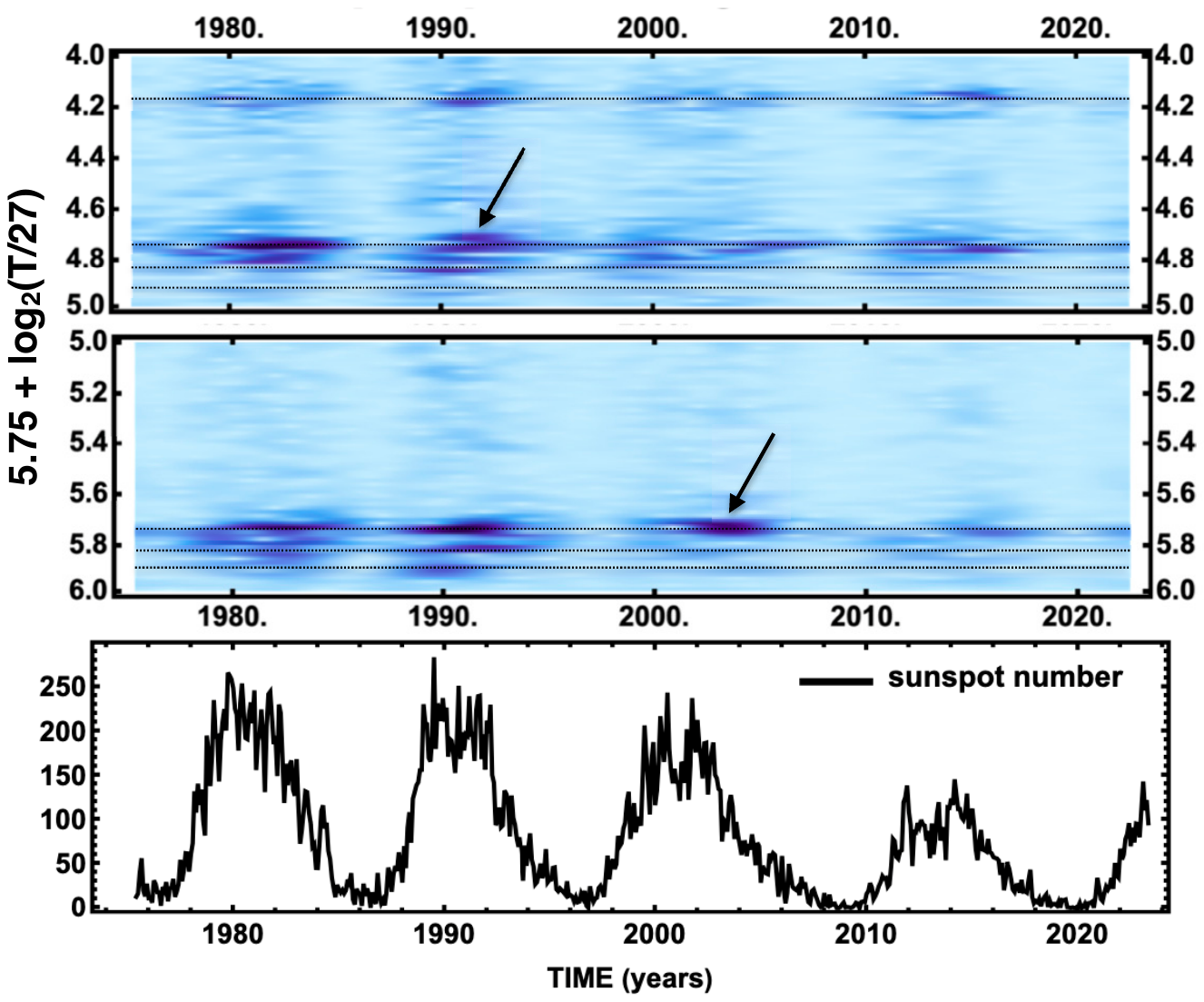}}}
\caption{Wavelet power in the region of the solar rotation period
(middle panel) and its second and third harmonics (top panel), displayed using
${\gamma}/2{\pi}=20$ and ${\gamma}/2{\pi}=40$, respectively.  Faint dotted lines
indicate periods of $T=27$ days, 28.5 days, and 30 days
near 5.8 on the vertical scale, and for their second harmonics near 4.8.  A single dotted
line indicates the third harmonic of $T=27$ days near 4.2.  The arrows indicate features
at 4.73 in 1990-1991 and 5.73 during 2002-2004, corresponding to rotation periods
${\sim}$26.5 days. 
(bottom) Monthly averaged sunspot number from the Royal Observatory of Belgium (SILSO).}
\label{fig:fig3}
\end{figure}
\noindent
been increased to 20 and 40 cycles per decay time, respectively, and in both
panels, the display resolution has been increased to  64 pixels per octave, corresponding
to $dT/T=0.022$ and $dT=0.27$ days.  For reference,
faint dotted lines are drawn at values of $T=27$ days, 28.5 days, and 30-days in the
middle panel, and at their second harmonics in the top panel.  Only the line
corresponding to $T=9$ days is shown for the third harmonic in the top panel. 
As before, the bottom panel shows the monthly averaged sunspot number from the
Royal Observatory of Belgium plotted for cycles 21-24 and the rising phase of cycle 25.

In Figure~3, the rotational fine structure is clearly visible in the dipole and quadrupole
modes, and to a lesser extent in the hexapole mode.  As found previously
\citep{Sheeley_2022}, the splitting is mainly one-sided, ranging from the equatorial
rotation period of about 27 days to 30 days, as one might expect for magnetic patterns
that are subject to differential rotation at latitudes between the equator and 45$^{\circ}$
\citep{NN_1951,SNOD_1983,SDV_1986}.  In addition, meridional flow and the poleward
component of supergranular diffusion ought to affect the rotation by making its latitudinal 
profile more rigid  \citep{SNW_1987,DEV_87,WANG_1998}.  The strong 30-day period is
visible for the dipole field in 1989-1990, and there is some weak 27-day power in
2022 at the start of the rising phase of sunspot cycle 25.  As noted previously
\citep{Sheeley_2022}, the fine structures in one sectoral mode are not reproduced in the
other sectoral modes.  These fine structures refer to different multipole patterns of field, and
are not the same structures rotating at different rates. 

A puzzling feature of this map is the presence of power at 4.73 on the vertical scale in the years 1990-1991.  This corresponds to a magnetic quadrupole rotating with a period of 13.24 days.
Interpreted as a second harmonic, this would imply that the fundamental is recurring with a
period of ${\sim}26.48$ days, which is noticeably smaller than the equatorial period of
26.9-27 days.  There is a similar feature near 5.73 in the middle panel during 2002-2004,
corresponding to a magnetic dipole field rotating with period of 26.5 days, which is again
noticeably smaller than the equatorial rotation period of the Sun's surface.  How can these
magnetic features drift in longitude faster than the equatorial surface rotates?  By means of
waves, as I will explain in the next section of this paper. 

\section{Summary and Discussion}
The purpose of this paper was to see if a wavelet analysis would reveal the fine structure
within each of the harmonic components of the mean field and show how that fine structure
varied with time during past sunspot cycles.  These objectives were achieved, but, to resolve
the rotational fine structures, it was necessary to increase ${\gamma}/2{\pi}$, the number
of waves per decay time of the wavelet, well beyond the value of approximately 1 that is
customarily used in the conventional Morlet transform.  This was illustrated in Figures~1-3,
which showed the distribution of wavelet power when ${\gamma}/2{\pi}$ ranged from 0.85
to 40.

The middle and upper panels of Figure~3 showed the rotational fine
structure in the 2-sector, 4-sector, and 6-sector fields, expressed as a function of time
during sunspot cycles 21-24 and early in the rising phase of cycle 25.  In addition
to power with the 27-day equatorial rotation period during each sunspot cycle, a substantial
amount of 2-sector power was visible with a 30-day rotation period during 1989-1990, reproducing
the result obtained previously using Fourier transforms \citep{Sheeley_2022}.

Figure~3 showed puzzling examples of apparent solar rotation with periods of
approximately 26.5 days, noticeably lower than the 26.9-27 day equatorial rotation period
of the Sun's surface.  In an attempt to understand this behavior, I examined a time-lapse movie
constructed from MWO Carrington maps of Ca II K-line emission during the years 1915-1979 and
from NSO Carrington maps of the photospheric magnetic field during 1980-2002.  This movie
was made previously as part of the work for a paper on the long-term variation of K-line emission  \citep{SCA_2011}.  Because these Carrington maps were displayed with the 27.27-day Carrington cadence, magnetic patterns poleward of about 15$^{\circ}$ (where the synodic rotation
period is about 27.27 days) moved rapidly eastward, and features equatorward of about 15$^{\circ}$ drifted westward.  The visual appearance was dramatic; high-latitude regions of each new sunspot
cycle seemed to `fly' off to the left, and the flux diffusing away from the last equatorial active
regions of each cycle drifted very slowly to the right.

In addition to these obvious indications of differential rotation, there were some flux
concentrations that seemed to move more rapidly to the west.  Their speeds were about twice
the drift speed of long-lived active regions at the Sun's equator, which is consistent with a
rotation period  ${\sim}$26.5 days.  (A period of 26.5 days corresponds to a rotation rate of
13.58$^{\circ}$day$^{-1}$, or 0.38$^{\circ}$day$^{-1}$ faster than the Carrington rate
of 13.20$^{\circ}$day$^{-1}$.  By comparison, the 26.9-day equatorial rate is
13.38$^{\circ}$day$^{-1}$, or 0.18$^{\circ}$day$^{-1}$ faster than the Carrington rate.
Thus, in the Carrington maps, the 26.5-day feature moves westward at a rate that is
0.38/0.18 = 2.1 times faster than flux drifting at the equatorial rate.)

However, a close examination of the movie showed that these fast westward motions were
not due to the differential rotation of long-lived magnetic regions.  Rather, they were
waves - apparent motions caused by the emergence of new active regions at  progressively
more western longitudes in the Carrington frame.  Similar wavelike motions have been
observed previously in Carrington stackplots of the NSO magnetic measurements by \cite{GAIHHZ_2015}.  In fact, their oxymoron, `simple complex IIIa', is
a very clear example of this fast wavelike motion in the N10-40$^{\circ}$ latitude band during
Carrington rotations 1656-1659 (14 June - 03 September 1977).  My measurements
of their Figure~2 give an effective rotation period of 26.3 days, compared to their
value of 26.5 days.  But the essential point is that the period was significantly
smaller than the 26.9-27.0-day equatorial rotation period of the Sun.  Consequently, the
motion cannot be caused by longitudinal transport of flux by solar rotation, at least
at the Sun's surface.

It is interesting that the 26.5-day period obtained from our wavelet measurements is nearly
the same as the ${\sim}$26.4-day equatorial rotation period found at a depth of
0.06$R_{\odot}$ (${\sim}$42 Mm) using GONG helioseismology measurements
\citep{2009LRSP....6....1H}.  Thus, the fast wavelike motion in Figure~3 occurs as if the
flux were emerging from a fixed longitude at that depth.  In terms of the popular word, `nesting'
which \cite{CZVNEST_1986} introduced for repeated eruptions of sunspot activity at the
same location (see also the reviews by \cite{vanD_2015} and by \cite{HATHAWAY_2015}),
we could call this fast wavelike motion `subsurface nesting'. 

This raises the question of whether all of the enhanced wavelet patterns of the mean
field are due to nesting.  We know that the long-lived recurrent patterns of 28-30-day
periods owe their existence to the continued emergence of bipolar regions with their
polarities in phase.  Otherwise, differential rotation winds up their flux and leaves a
weaker field rotating with the 27-day period of the unwound flux at the
equator.  This is consistent with our more recent demonstration that the mean field
is dominated
by the horizontal dipole and quadrupole fields (and to a lesser extent the hexapole field)
\citep{Sheeley_2022}.  So, the active regions must emerge in a way that reinforces the
lower-order multipoles of the non-axisymmetric field.  It does not matter whether this
systematic emergence occurs in active longitudes as described by \cite{GAIHHZ_2015},
or whether it occurs by chance as in the randomized doublets of \cite{SDV_1986}.  More
recently, \cite{Hudson_2014} have discussed this effect as a way of reinforcing long-lived
polarity patterns at Hale sector boundaries.

It is interesting that when \cite{SDV_1986} removed supergranular diffusion and meridional
flow from their simulations during the declining phase of the sunspot cycle (when all of the
sources emerged below 20$^{\circ}$ latitude), the 28-29-day recurrence patterns disappeared.
In other words, a poleward component of transport was essential for obtaining the 28-29-day
periods, and with this transport, the slowly rotating patterns could be obtained from the
flux in low-latitude active regions (providing that they continued to emerge in phase.

Also, this is consistent with the results of \cite{SNW_1987}, who found that the
large-scale field
rotates differentially when the source rate is high, but begins to weaken and rotate rigidly
with the 27-day equatorial period when the flux stops emerging and the surviving
non-equatorial flux migrates poleward and is eliminated by differential rotation and
supergranular diffusion.  Thus, it seems likely that spectrally resolved wavelet patterns
correspond to active-region nests at different latitudes and with strong horizontal dipole
and/or quadrupole moments.  Wavelet maps like those in Figure~3 indicate the
sunspot-cycle distribution of those nests, and ${\gamma}$-sequences, like those in
Figure~2, provide a way of finding the individual sources that maintain the strengths of
these patterns. 

I am grateful to Kristopher Klein (LPL/UA) for suggesting wavelet transforms as a way of tracking
solar features of different rotational periods.  Also, I am grateful to Chen Shi (UCLA) for providing
details of how he made wavelet maps of the radial component of the interplanetary magnetic field
observed at the Parker Solar Probe.  Wilcox Solar Observatory data used in this study were
obtained \textit{via} the web site http://wso.stanford.edu courtesy of J.T. Hoeksema. NSO data
were acquired by SOLIS instruments operated by NISP/NSO/AURA/NSF.  Sunspot numbers were obtained from WDC-SILSO, Royal Observatory of Belgium, Brussels.

\appendix
\section{Relations between ${\omega}$ and ${\omega}_{0}$}
As described in Sections 2.1 and 2.2 of the text, the basic concepts of wavelets and wavelet
transforms are relatively straightforward.  The most difficultly I had was in describing the quantity that was plotted on the vertical axis of the map of wavelet power.  In principle, it ought to be the logarithm
of the wavelet scale, $\log_{2}s$.  But in practice, what I really wanted was a logarithmic measure of of the oscillation frequency, ${\omega}$, or equivalently, its corresponding period, $T = 2{\pi}/{\omega}$.  Consequently, I needed to find a suitable relation between wavelet scale and period similar to what others have obtained in the past using a different notation for the wavelet frequency and a more limited
range of ${\gamma}/2{\pi}$ \citep{BOB_2002,POD_2009,TOR_98,SHI_2022,MATH_2023}.  I proceeded
as follows:
 
First, I note that the Gaussian decay factor, $e^{-(1/2)(t/s)^2}$, does not change the frequency
of the wavelet because its Fourier transform,
$\hat{\psi}({\omega})$ varies as $e^{-(1/2)({\gamma}-{\omega}s)^2}$, which peaks at ${\omega}={\gamma}/s={\omega}_{0}$.  However, to be admissible as a wavelet,
${\psi}$ must have a mean value of zero \citep{FAR_92,TOR_98}, which can be achieved by subtracting a constant term from the oscillating factor, and then adjusting that term so that
$\int_{-\infty}^{\infty}{\psi}(t,s)dt=0$.  In this case, the constant term is $e^{-{\gamma}^2/2}$,
and the shifted wavelet becomes
\begin{equation}
{\psi} \left (\frac{t-{\tau}}{s} \right )~{\sim}~e^{-\frac{1}{2}((t-{\tau})/s)^2}~
\left [ e^{i{\gamma} ((t-{\tau})/s)}~-~e^{-{\gamma}^2/2} \right ].
\end{equation}
Now, the Fourier transform changes to
\begin{equation}
\hat{\psi}({\omega})~=~sA{\sqrt{2\pi}} \left[ e^{-\frac{1}{2}({\gamma}-{\omega}s)^2}~-~ 
e^{-{\gamma}^2/2}~e^{-\frac{1}{2}({\omega}s)^2} \right ],
\end{equation}
where $A={\pi}^{-1/4} s^{-1/2} [1~+~e^{-{\gamma}^2}~-~2e^{-(3/4){\gamma}^2} ]^{-1/2}$
is a normalization constant chosen so that $\int_{-\infty}^{\infty}{\psi}{\psi}^{*}dt = 1$.
With this change, the location of the peak of $\hat{\psi}({\omega})$ changes, and it is necessary
to find its new location by computing ${\partial}{\hat{\psi}}/{\partial}{\omega}$ and setting it
equal to 0.  This gives the relation
\begin{equation}
e^{{{\gamma}^2}x}~=~\frac{x}{x-1}
\end{equation}
where $x={\omega}/{\omega}_{0}$.  For the exponential to become large, the value of $x$ on
the right hand side of this equation must approach 1.  Consequently, we can set $x=1+{\epsilon}$
and solve for the small quantity, $\epsilon$.  The result is ${\epsilon}~{\approx}~e^{-{\gamma}^2}$.
Therefore,
\begin{equation}
x~=~\frac{{\omega}}{{\omega}_{0}}~{\approx}~1+e^{-{\gamma}^2},
\end{equation}
and the correction is
less than 1\% for ${\gamma}~{\geq}~2.14$.  We can find a second-order correction, ${\delta}$, by
defining $x_{1}=1+e^{-{\gamma}^2}$, and substituting $x=x_{1}-{\delta}$ into Eq(A3).  After some algebra, we obtain ${\delta}=x_{1}-\{1-e^{-{\gamma}^2 x_{1}} \}^{-1}$, and $x=\{1-e^{-{\gamma}^2 x_{1}} \}^{-1}$.  This second-order solution lies within 2\% of the numerical solution of Eq(A3) for
${\gamma}=1$, and rapidly agrees to better than 1\% as ${\gamma}$ exceeds 1.25.

So the frequency of the wavelet is very close to ${\omega}_{0}$.  What about the frequency
obtained from the wavelet transform?  To find out, I will compute the wavelet power $|F(t,s)|^2$
for a time series of angular frequency, ${\omega}$, given by $\cos{\omega}t$, and look for the value
of $s$ that makes this power a maximum.  This will provide another relation between ${\omega}$
and $s$, that can be combined with Eq(4) to relate
${\omega}$ and ${\omega}_{0}$.   For this purpose, we return to
 Eq(6) where we set $B_{m}(t)=\cos{{\omega}t}$ and use the value of ${\psi}$ given by Eq(A1):
 \begin{equation}
F(t,s)~=~\frac{1}{\sqrt{s}}\int_{-\infty}^{+\infty}\cos{\omega}t~{\psi}^{*}(\frac{t-{\tau}}{s})d{\tau}~=~
\frac{{\pi}^{-1/4}}{\sqrt{s}} \int_{-\infty}^{+\infty}\cos{\omega}t~e^{-\frac{1}{2}\{(t-{\tau})/s\}^2}[
e^{-i{\gamma}\{(t-{\tau})/s\}}~-~e^{-{\gamma}^2/2}]d{\tau}.
\end{equation}
 
 Evaluating this integral, we obtain
 \begin{equation}
 F(t,s)~{\sim}~s^{1/2}e^{-{\gamma}^2/2} \left [ \cos{\omega}t \{\cosh({\gamma}{\omega}s)-1 \}~-~
 i~ \sin{\omega}t \sinh({\gamma}{\omega}s) \right ],
 \end{equation}
where constant factors depending on ${\pi}$ have been dropped.  Next, because $F(t,s)$ is
a complex number, we multiply $F(t,s)$ by its complex conjugate to obtain
\begin{equation}
|F(t,s)|^2~{\sim}~s e^{-{\gamma}^2}~ [\cosh({\gamma}{\omega}s)-1 ]~
[{\cosh({\gamma}\omega}s)-\cos2{\omega}t ].
\end{equation}
At this point, we recognize that ${\omega}s~{\gtrsim}~{\gamma}$, in which case
$\cosh({\gamma}{\omega}s)~{\gtrsim}~\cosh({\gamma}^2) >> 1$.  Consequently, Eq(A7)
reduces to
\begin{equation}
|F(t,s)|^2~{\sim}~s e^{-{\gamma}^2} \cosh^{2}({\gamma}{\omega}s),
\end{equation}
and $|F(t,s)|$ becomes
\begin{equation}
|F(t,s)|~{\sim}~s^{1/2} e^{{-{\gamma}^2}/2} \cosh({\gamma}{\omega}s).
\end{equation}
It is interesting to note that if we had ignored the $e^{{-{\gamma}^2}/2}$ term in
Eq(A1) from the beginning, we would have obtained
\begin{equation}
|F(t,s)|^2~{\sim}~s e^{-{\gamma}^2} e^{-({\omega}s)^2}~ [\cosh^{2}({\gamma}{\omega}s)-\cos^{2}{\omega}t ]
\end{equation}
instead of Eq(A7).  Then, the condition that $\cosh^{2}({\gamma}{\omega}s)>>\cos^{2}{\omega}t$
would have led to Eq(A8) and Eq(A9).  This means that the approximation that we used to get
there the first time is equivalent to neglecting the extra term in the wavelet that arose when we
forced $\int_{-\infty}^{\infty}{\psi}dt = 0$.  So, looking ahead, whatever we find from Eq(A9)
will be produced by the original wavelet without the extra $e^{{-{\gamma}^2}/2}$ term.

Now, let us use Eq(A9) to evaluate ${\partial}|F(t,s)|/{\partial}{s}$ and then set it equal to 0 to find
the value of $s$ that makes $|F(t,s)|$ a maximum.  The derivative is
\begin{equation}
\frac{{\partial}|F(t,s)|}{{\partial}s}~=~\frac{e^{{-({\omega}s)^2}/2}}{2{\sqrt{s}}}
 \left [- \{2({\omega}s)^2-1\}\cosh({\gamma}{\omega}s)+
 2({\gamma}{\omega}s)\sinh({\gamma}{\omega}s) \right ],
\end{equation}
and the resulting equation for $s$ is
\begin{equation}
\tanh({\gamma}{\omega}s)~=~\frac{2({\omega}s)^2-1}{2({\gamma}{\omega}s)}.
\end{equation}
Substituting $s={\gamma}/{\omega}_{0}$ from Eq(4), and converting from
the hyperbolic tangent to ordinary exponentials, we obtain
\begin{equation}
e^{2{\gamma}^2 x}~=~\frac{x+x^2-\frac{1}{2{\gamma}^2}}{x-x^2+\frac{1}{2{\gamma}^2}},
\end{equation}
where $x={\omega}/{\omega}_{0}$.  Eq(A13) is analogous to Eq(A3), and can be solved the same
way.  As before, we assume that the denominator of the right hand side is close to 0, and solve
for the positive root of $x-x^2+(1/2{\gamma}^2)=0$.  The result is 
\begin{equation}
x~=~\frac{{\omega}}{{\omega}_{0}}~=~(1/2) \left  [ 1+{\sqrt{1+\frac{2}{{\gamma}^2}}}  \right ],
 \end{equation}
 consistent with Eq(B8) of \cite{POD_2009} using slightly different notation.
 
 If we regard Eq(A14) as the first-order solution, $x_{1}$, then we can write $x=x_{1}-{\epsilon}$
 and substitute it into Eq(A13) to obtain
 ${\epsilon}=\left [2x_{1}/(2x_{1}-1) \right ]~e^{-2{\gamma}^2 x_{1}}$ as the second-order
 correction.  In this case, the second-order solution for $x$ becomes
 \begin{equation}
 x~=~\frac{{\omega}}{{\omega}_{0}}~=~x_{1}-\left (\frac{2x_{1}}{2x_{1}-1} \right ) e^{-2{\gamma}^2 x_{1}},
 \end{equation}
 where $x_{1}$ is the first-order solution given by Eq(A14).  Thus, for ${\gamma} ~{\gtrsim}~6$,
 this second-order correction is negligible and the first-order solution, given by Eq(A14),
 is accurate to about 1\%.  There is no difference between ${\omega}$
 and ${\omega}_{0}$ for the larger values of ${\gamma}$ that resolve the 27-day, 28.5-day,
 and 30-day components of differential rotation.   For smaller values of ${\gamma}/2{\pi}$,
 one can use Eqs(A14) and (A15).
 
 \section{The Transition Between High Spectral and Spatial Resolution}
Regardless of the values of $s$ and ${\omega}_{0}$ (or ${\gamma}$), the
wavelet and its Fourier transform satisfy the Heisenberg uncertainty principle in the form
${\Delta}{\omega}{\Delta}t=1/2$, provided that ${\Delta}{\omega}$ and ${\Delta}t$ are both
defined as the root-mean-square differences from their average values.  As mentioned
below, this provides a basis for understanding the transition between
the maps of high spectral resolution and high spatial resolution like those shown
in Figure~2.

Let's begin with ${\psi}(t)$ and its Fourier transform, $\hat{\psi}({\omega})$: 
\begin{subequations}
\begin{align}
 {\psi}(t)~=~e^{-\frac{1}{2}(t/s)^2}e^{i{\omega}_{0}t},\\
 \hat{\psi}({\omega})~=~\int_{-{\infty}}^{{\infty}}
 {\psi}(t)e^{-i{\omega}t}dt~=~{\sqrt2}s{\sqrt{\pi}} e^{-({s^2}/2)({\omega}-{\omega}_{0})^2}.
 \end{align}
\end{subequations}
In this case, the mean squared averages are
\begin{subequations}
\begin{align}
<({\Delta}t)^2>~=~\frac{\int_{-{\infty}}^{{\infty}} {t^2}{\psi}(t){\psi}^{*}(t) dt}
{\int_{-{\infty}}^{{\infty}}{\psi}(t){\psi}^{*}(t) dt}~=~\frac{s^2}{2},\\
<({\Delta}{\omega})^2>~=~\frac{\int_{-{\infty}}^{{\infty}}{({\omega}-{\omega}_{0})^{2} \hat{\psi}({\omega})\hat{\psi}^{*}({\omega})d{\omega}}}
{\int_{-{\infty}}^{{\infty}}{ \hat{\psi}({\omega})\hat{\psi}^{*}({\omega})d{\omega}}}~=~\frac{1}{2s^2}.
 \end{align}
\end{subequations}
Consequently, the root-mean-square values are 
\begin{subequations}
\begin{align}
({\Delta}t)_{rms}~=~\frac{s}{\sqrt{2}},\\
({\Delta}{\omega})_{rms}~=~\frac{1}{s{\sqrt{2}}},
\end{align}
\end{subequations}
and their product is
\begin{equation}
({\Delta}{\omega})_{rms}({\Delta}t)_{rms}~=~\frac{1}{2},
\end{equation}
independent of $s$.  Also, if the energy, $E$, of the wave packet were $E={\hbar}{\omega}$,
then
\begin{equation}
({\Delta}{E})_{rms}({\Delta}t)_{rms}~=~\frac{\hbar}{2},
\end{equation}
which is the conventional form of the Heisenberg uncertainty principle.

As mentioned in the text, the spectral resolution varies as $1/s$, so that Eq(B18b) gives the spectral resolution in terms of the scale, $s$.  Likewise,  Eq(B18a) gives the temporal resolution in terms
of $s$.  When combined with Eq(9) of Section 3 ($s = ({\gamma}/2{\pi})T$), these equations provide a way of calculating these resolutions as a function of ${\gamma}/2{\pi}$, and therefore of tracking the evolution of wavelet power from high temporal resolution to high spectral resolution as was done in Figure~2.

Let's pursue this matter further by recalling from Figure~2 that well-defined frequencies in
the high spectral resolution maps (with ${\gamma}/2{\pi}~{\sim}~8-16$) typically correspond to
bursts of short-lived features in the high temporal resolution maps
(with ${\gamma}/2{\pi}~{\sim}~1$).  In particular, the quadrupole feature at 4.73 on the vertical axis
during 1990-1991 (also marked by an arrow in Figure~3) corresponds to
a 0.5-yr burst of 2-3 short-lived features in the high temporal resolution maps at the top of Figure~2.
Other bursts are visible with lifetimes ranging from 0.4 yr to about 1 yr.  The closely spaced
doublet at 4.73-4.75 during 2002-2004 (also marked by an arrow in Figure~3) corresponds to one
of the strongest and longest-lived bursts in the high spatial resolution maps.  And, as mentioned in Section 3, the 30-day 2-sector pattern in 1989-1990 seems to have originated in separate
bursts in April 1989 and January 1990.

It is tempting to wonder if some of these temporal components are produced by the
interference between closely spaced frequencies in the high spectral resolution maps, analogous
to the periodic stripes that occur in Bartels displays of interplanetary magnetic field polarity when long-lived patterns of 27 days and 28.5 days overlap in time \citep{SVAL_1975,WS_1994}.  However, the resulting `beat' frequencies correspond to periods that are somewhat larger than the durations of these bursts.  In fact, 27-day and 28.5-day features give a period of 27${\times}$28.5/1.5 = 513 days = 1.4 yr, and the other frequency pairs give even larger periods.  This does not include the damping associated with the wavelet scale, $s$, and additional work is necessary to resolve this issue with confidence.  Meanwhile, I suppose that these components of temporal fine structure are enhancements of the Sun's equatorial dipole, quadrupole, and hexapole moments caused by the ongoing emergence of active regions in `nests' rotating with 27-day, 28.5-day, and other periods. 

 \section{The Mean-Field of a Photospheric Doublet}
Having arrived at the idea that the mean field is due to the suitable juxtaposition of magnetic
doublets, I am interested to know what the contribution of a single doublet is, and the
circumstances under which it could appreciably affect the mean field.  An idealized magnetic
doublet can be represented by an expression of the form
\begin{equation}
B_{r}~=~\frac{{\Phi}_{0}}{R^2} \left [\frac{{\delta}({\phi}-{\phi}_{L}){\delta}({\theta}-{\theta}_{L})}{\sin{\theta}_{L}}~-~ \frac{{\delta}({\phi}-{\phi}_{F}){\delta}({\theta}-{\theta}_{F})}{\sin{\theta}_{F}}                     \right ],
\end{equation}
where $L$ and $F$ refer to the leader and follower poles of the doublet, respectively, and the delta
functions indicate the concentrated nature of those poles.  It is easy
to confirm that this magnetic doublet satisfies
\begin{equation}
\int_{0}^{\pi}\int_{0}^{2{\pi}}B_{r}({\theta},{\phi})R^2 \sin{\theta}d{\theta}d{\phi}~=~{\Phi}_{0}-{\Phi}_{0}~=~0.
\end{equation}
To obtain the mean field of this doublet, we simply evaluate the integral
\begin{equation}
B_{m}~=~\frac{1}{{\pi}}\int_{0}^{\pi}\int_{0}^{\pi}B_{r}({\theta},{\phi})(\sin{\theta}\cos{\phi})^2 \sin{\theta}d{\theta}d{\phi}.
\end{equation}
The result is
\begin{equation}
B_{m}~=~(\frac{{\Phi}_{0}}{{\pi}R^2})\left [ \sin^{2}{\theta}_{L}\cos^{2}{\phi}_{L}~-~\sin^{2}{\theta}_{F}\cos^{2}{\phi}_{F} \right ].
\end{equation}
Next, we define the separations between the leader and follower poles and their mid-points by
${\theta}_{L}={\theta}_{0}+{\Delta}{\theta}/2$,  ${\theta}_{F}={\theta}_{0}-{\Delta}{\theta}/2$,
${\phi}_{L}={\phi}_{0}+{\Delta}{\phi}/2$, and ${\phi}_{F}={\phi}_{0}-{\Delta}{\phi}/2$, where
${\Delta}{\theta}$ and ${\Delta}{\phi}$ are the pole separations and ${\theta}_{0}$ and
${\phi}_{0}$ are the mid-points between the respective poles.  Substituting these relations into
Eq(C24) and using the small-angle relations for ${\Delta}{\theta}$ and ${\Delta}{\phi}$, we obtain
\begin{equation}
B_{m}~{\approx}~(\frac{{\Phi}_{0}{\Delta}{\theta}}{{\pi}R^2})\cos^{2}{\phi}_{0} \sin2{\theta}_{0}~-~
(\frac{{\Phi}_{0}{\Delta}{\phi}}{{\pi}R^2})\sin^{2}{\theta}_{0}\sin2{\phi}_{0}.
\end{equation}
Now, let's evaluate the root-mean-square value of the expression in Eq(C25) using the relation

\noindent
$<B^{2}_{m}>~=~\int_{0}^{\pi}B^{2}_{m}d{\phi}_{0}/ \int_{0}^{\pi}d{\phi}_{0}$.  The result is
\begin{equation}
<B^{2}_{m}>~=~\frac{3}{8}(\frac{{\Phi}_{0}{\Delta}{\theta}}{{\pi}R^2})^{2}\sin^{2}2{\theta}_{0}~+~
\frac{1}{2}(\frac{{\Phi}_{0}{\Delta}{\phi}}{{\pi}R^2})^{2}\sin^{4}{\theta}_{0}.
\end{equation}
If we neglect ${\Delta}{\theta}$ compared to ${\Delta}{\phi}$ and take the square root,
we obtain
\begin{equation}
B^{rms}_{m}~=~\frac{1}{\sqrt2}(\frac{{\Phi}_{0}{\Delta}{\phi}}{{\pi}R^2})\sin^{2}{\theta}_{0}.
\end{equation}
Thus, the rms mean field of a doublet, located near the equator where ${\theta}_{0}={\pi}/2$,
is roughly equal to its doublet moment, ${\Phi}_{0} {\Delta}{\phi}$, divided by the area
of the visible disk, ${\pi}R^2$.  More accurately,
$B^{rms}_{m}~{\approx}~0.707({\Phi}_{0}{\Delta}{\phi}/{\pi}R^2)$.

If this idealized magnetic doublet has a pole strength ${\Phi}_{0} = 10{\times}10^{21}$Mx and a longitudinal pole separation ${\Delta}{\phi}~ {\sim}~10^{\circ}$ (${\sim}10^{5}$ km),  corresponding to
a mid-sized active region, then its contribution to the Sun's mean field would be only about 0.08 G.  This is an order of magnitude smaller than the 1 G peaks that typically occur when the Sun's equatorial dipole and quadrupole moments are strong.  To achieve these strong fields, it would take a nest of several large bipolar magnetic regions arranged with their polarities in phase, as we have found in spatially resolved observations around sunspot maximum and during the initial declining phase of the sunspot cycle \citep{Sheeley_2015}.

\section{The Open Flux of a Photospheric Doublet}
Here, we extend the previous calculation to determine how much open flux is contributed by the
idealized magnetic doublet given by Eq(C21).  To obtain this open flux, we assume a potential field
whose angular components, $B_{\theta}$ and $B_{\phi}$ vanish at a spherical source surface located
at a radial distance $R_{ss}$, and whose radial component, $B_{r}$ matches the radial component
of the doublet field given by Eq(C21) .  Eventually, we will select $R_{ss}=2.5R$, where R is the
solar radius, but, for the moment, let's consider a general value of $R_{ss}$.  Our objective is
to find the source surface field and then integrate its positive value over the source surface.

Rather than repeating the solution of this well-known boundary-value problem, I will take the
solution from Eq(4a) of a previous paper by  \cite{NASH_1988}.  Setting $r=R_{ss}$ in their
equation, the source-surface field becomes:
\begin{equation}
B_{ss}~=~B_{r}(R_{ss},{\theta},{\phi})~=~\sum_{l=0}^{\infty}\sum_{m=-l}^{l}
\left [ \frac{(2l+1)(R/R_{ss})^{l+2}}{l+1+l(R/R_{ss})^{2l+1}} \right ]c_{lm}Y^{m}_{l}({\theta},{\phi}),
\end{equation}
 where $c_{lm}$ are the spherical harmonic components of the doublet, given by
 \begin{equation}
 B_{ph}~=~\sum_{l=0}^{\infty}\sum_{m=-l}^{l}c_{lm}Y^{m}_{l}({\theta},{\phi}).
 \end{equation}
 When $B_{ph}$
 is given by the doublet field of Eq(C21) with ${\theta}_{L}={\theta}_{F}={\theta}$ and
 ${\phi}_{L}-{\phi}_{F}={\Delta}{\phi}<<{\pi}/2$, the solution to Eq(D29) is
 \begin{equation}
 c_{lm}~=~-imN_{lm}P^{m}_{l}(0) \left (\frac{{\Phi}_{0}{\Delta}{\phi}}{R^2} \right),
 \end{equation}
 where $i=\sqrt{-1}$, $N_{lm}=\sqrt{ \frac{2l+1}{4{\pi}}\frac{(l-m)!}{(l+m)!} }$, and $P^{m}_{l}(0)$
 is an Associated Legendre polynomial function of $x$ evaluated at $x=0$.
 
 Next, we suppose that the resulting field will be dominated by the contributions of the horizontal
 dipole and quadrupole, and limit the sum in Eq(D28) to terms of the form $Y^{{\pm}1}_{1}$ and
 $Y^{{\pm}2}_{2}$.  In this case, the source-surface field is
 \begin{equation}
 B_{ss}({\theta},{\phi})~=~a\sin{\theta}\sin{\phi}~+~b\sin^{2}{\theta}\sin{2{\phi}},
 \end{equation}
 where
 \begin{equation}
 a~=~\frac{3}{4} \left [ \frac{3(R/R_{ss})^{3}}{2+(R/R_{ss})^{3}} \right ]
  \left (\frac{{\Phi}_{0} {\Delta}{\phi}}{{\pi}R^{2}} \right )~{\approx}~\frac{9}{8}(\frac{R}{R_{ss}})^{3}
   \left (\frac{{\Phi}_{0} {\Delta}{\phi}}{{\pi}R^{2}} \right )
 \end{equation}
 and
 \begin{equation}
 b~=~\frac{15}{8} \left [ \frac{5(R/R_{ss})^{4}}{3+2(R/R_{ss})^{5}} \right ]
  \left (\frac{{\Phi}_{0} {\Delta}{\phi}}{{\pi}R^{2}} \right )~{\approx}~\frac{25}{8}(\frac{R}{R_{ss}})^{4}
   \left (\frac{{\Phi}_{0} {\Delta}{\phi}}{{\pi}R^{2}} \right ).
 \end{equation}
 It is interesting that $b/a~{\approx}~(25/9)(R/R_{ss})$, which is approximately 1.11 when $R_{ss}/R$
 has the conventional value of 2.5.  Thus, $a~{\sim}~b$, so that the horizontal dipole and quadrupole
 probably make comparable contributions to the open flux.  In fact, we can confirm this by evaluating their separate integrals over the regions of positive flux:
 \begin{equation}
 {\Phi}_{open}~{\approx}~\frac{9}{8{\pi}}(\frac{R}{R_{ss}})
 \int_{0}^{{\pi}}\sin^{2}{\theta}\sin{\theta}d{\theta}
 \int_{0}^{{\pi}}\sin{\phi}d{\phi}~({{\Phi}_{0}{\Delta}{\phi}})~=~
 \frac{9}{8}(\frac{R}{R_{ss}})({{\Phi}_{0}{\Delta}{\phi}})~=~0.45~({{\Phi}_{0}{\Delta}{\phi}})
 \end{equation}
 for the dipole field, and
 \begin{equation}
  {\Phi}_{open}~{\approx}~(\frac{25}{8{\pi}})(\frac{R}{R_{ss}})^{2}
  \int_{0}^{{\pi}}\sin^{2}{\theta}d{\theta}
  ~~2\int_{0}^{{\pi/2}}\sin{2{\phi}}d{\phi}~
  ({{\Phi}_{0}{\Delta}{\phi}})~=~(\frac{25}{3{\pi}})(\frac{R}{R_{ss}})^{2}~({{\Phi}_{0}{\Delta}{\phi}})
  ~=~0.42~({{\Phi}_{0}{\Delta}{\phi}})
 \end{equation}
 for the quadrupole field.  The extra factor of 2 in Eq(D35) allows for the fact that the quadrupole has 2 separate quadrants of positive field.  The result that
${\Phi}_{open}=0.45({\Phi}_{0}{\Delta}{\phi})$ in Eq(D34) agrees with Eq(1) of \cite{Sheeley_2015},
and its derivation in Eqs(D28)-(D34) provides the documentation for the steps that were only outlined
there. 
 
 When both harmonics are present at the same time, we can obtain the positive flux by numerically
integrating the absolute value of the field over the entire sphere and dividing the result by 2.  Schematically, the flux is given by
 \begin{equation}
 {\Phi}_{open}~=~\frac{1}{2}\int_{0}^{{\pi}}\int_{0}^{2{\pi}} |p \sin{\theta}\sin{\phi}+q\sin^{2}{\theta}\sin{2{\phi}}|\sin{\theta}d{\theta}d{\phi}~ ({\Phi}_{0}{\Delta}{\phi}) ,
 \end{equation}
 where $p=aR^{2}_{ss}/({\Phi}_{0}{\Delta}{\phi})~{\approx}~(9/8{\pi})(R/R_{ss})=0.143$ and
 $q=bR^{2}_{ss}/({\Phi}_{0}{\Delta}{\phi})~{\approx}~(25/8{\pi})(R/R_{ss})^2=0.159$.  The result is
 \begin{equation}
  {\Phi}_{open}~=~0.55~({{\Phi}_{0}{\Delta}{\phi}}).
 \end{equation}
So when both the horizontal dipole and quadrupole are included, the amount of open flux is
approximately $0.55({\Phi}_{0}{\Delta}{\phi})$ where the approximation was obtained by neglecting
terms of order $(R/R_{ss})^3=(0.4)^{3}=0.064$ in the expression for $a$ and $(0.4)^{5}=0.010$
in $b$.  Without this approximation, the combined open flux drops slightly to 0.54 of the doublet moment, mainly due to a drop of the dipole contribution from 0.45 to 0.44.  The essential point
is that for a current-free corona, the open flux of an idealized doublet located near
the equator is about half of its doublet moment.  This is comparable to
$0.71 ({\Phi}_{0}{\Delta}{\phi})$,  the flux obtained from Eq(C27) by multiplying the rms
mean field by the area of the visible solar disk.   Like the mean field, several suitably
arranged doublets of modest size will be required to contribute appreciably to the amount of
open flux.









\bibliography{mwave}{}

\begin{thebibliography}{}
\expandafter\ifx\csname natexlab\endcsname\relax\def\natexlab#1{#1}\fi
\providecommand{\url}[1]{\href{#1}{#1}}
\providecommand{\dodoi}[1]{doi:~\href{http://doi.org/#1}{\nolinkurl{#1}}}
\providecommand{\doeprint}[1]{\href{http://ascl.net/#1}{\nolinkurl{http://ascl.net/#1}}}
\providecommand{\doarXiv}[1]{\href{https://arxiv.org/abs/#1}{\nolinkurl{https://arxiv.org/abs/#1}}}

\bibitem[{Boberg {et~al.}(2002)Boberg, Lundstedt, Hoeksema, Scherrer, \&
  Liu}]{BOB_2002}
Boberg, F., Lundstedt, H., Hoeksema, J.~T., Scherrer, P.~H., \& Liu, W. 2002,
  Journal of Geophysical Research: Space Physics, 107, SSH 15,
  \dodoi{https://doi.org/10.1029/2001JA009195}

\bibitem[{{Castenmiller} {et~al.}(1986){Castenmiller}, {Zwaan}, \& {van der
  Zalm}}]{CZVNEST_1986}
{Castenmiller}, M.~J.~M., {Zwaan}, C., \& {van der Zalm}, E.~B.~J. 1986,
  \solphys, 105, 237, \dodoi{10.1007/BF00172045}

\bibitem[{{DeVore}(1987)}]{DEV_87}
{DeVore}, C.~R. 1987, \solphys, 112, 17, \dodoi{10.1007/BF00148484}

\bibitem[{{Farge}(1992)}]{FAR_92}
{Farge}, M. 1992, Annual Review of Fluid Mechanics, 24, 395,
  \dodoi{10.1146/annurev.fl.24.010192.002143}

\bibitem[{{Gaizauskas} {et~al.}(1983){Gaizauskas}, {Harvey}, {Harvey}, \&
  {Zwaan}}]{GAIHHZ_2015}
{Gaizauskas}, V., {Harvey}, K.~L., {Harvey}, J.~W., \& {Zwaan}, C. 1983, \apj,
  265, 1056, \dodoi{10.1086/160747}

\bibitem[{{Hathaway}(2015)}]{HATHAWAY_2015}
{Hathaway}, D.~H. 2015, Living Reviews in Solar Physics, 12, 4,
  \dodoi{10.1007/lrsp-2015-4}

\bibitem[{{Howe}(2009)}]{2009LRSP....6....1H}
{Howe}, R. 2009, Living Reviews in Solar Physics, 6, 1,
  \dodoi{10.12942/lrsp-2009-1}

\bibitem[{{Hudson} {et~al.}(2014){Hudson}, {Svalgaard}, \&
  {Hannah}}]{Hudson_2014}
{Hudson}, H.~S., {Svalgaard}, L., \& {Hannah}, I.~G. 2014, \ssr, 186, 17,
  \dodoi{10.1007/s11214-014-0121-z}

\bibitem[{{Nash} {et~al.}(1988){Nash}, {Sheeley}, \& {Wang}}]{NASH_1988}
{Nash}, A.~G., {Sheeley}, N.~R., J., \& {Wang}, Y.~M. 1988, \solphys, 117, 359,
  \dodoi{10.1007/BF00147253}

\bibitem[{{Newton} \& {Nunn}(1951)}]{NN_1951}
{Newton}, H.~W., \& {Nunn}, M.~L. 1951, \mnras, 111, 413,
  \dodoi{10.1093/mnras/111.4.413}

\bibitem[{{Podesta}(2009)}]{POD_2009}
{Podesta}, J.~J. 2009, \apj, 698, 986, \dodoi{10.1088/0004-637X/698/2/986}

\bibitem[{{Sheeley} {et~al.}(2011){Sheeley}, {Cooper}, \&
  {Anderson}}]{SCA_2011}
{Sheeley}, N.~R., J., {Cooper}, T.~J., \& {Anderson}, J.~R.~L. 2011, \apj, 730,
  51, \dodoi{10.1088/0004-637X/730/1/51}

\bibitem[{{Sheeley} \& {DeVore}(1986)}]{SDV_1986}
{Sheeley}, N.~R., J., \& {DeVore}, C.~R. 1986, \solphys, 104, 425,
  \dodoi{10.1007/BF00159092}

\bibitem[{{Sheeley} {et~al.}(1985){Sheeley}, {DeVore}, \& {Boris}}]{SHEDB_1985}
{Sheeley}, N.~R., J., {DeVore}, C.~R., \& {Boris}, J.~P. 1985, \solphys, 98,
  219, \dodoi{10.1007/BF00152457}

\bibitem[{{Sheeley} {et~al.}(1987){Sheeley}, {Nash}, \& {Wang}}]{SNW_1987}
{Sheeley}, N.~R., J., {Nash}, A.~G., \& {Wang}, Y.~M. 1987, \apj, 319, 481,
  \dodoi{10.1086/165472}

\bibitem[{{Sheeley} \& {Wang}(2015)}]{Sheeley_2015}
{Sheeley}, N.~R., J., \& {Wang}, Y.~M. 2015, \apj, 809, 113,
  \dodoi{10.1088/0004-637X/809/2/113}

\bibitem[{Sheeley(2022)}]{Sheeley_2022}
Sheeley, N.~R. 2022, The Astrophysical Journal, 937, 87,
  \dodoi{10.3847/1538-4357/ac86d6}

\bibitem[{{Shi} {et~al.}(2022){Shi}, {Panasenco}, {Velli}, {Tenerani},
  {Verniero}, {Sioulas}, {Huang}, {Brosius}, {Bale}, {Klein}, {Kasper}, {de
  Wit}, {Goetz}, {Harvey}, {MacDowall}, {Malaspina}, {Pulupa}, {Larson},
  {Livi}, {Case}, \& {Stevens}}]{SHI_2022}
{Shi}, C., {Panasenco}, O., {Velli}, M., {et~al.} 2022, \apj, 934, 152,
  \dodoi{10.3847/1538-4357/ac7c11}

\bibitem[{{Snodgrass}(1983)}]{SNOD_1983}
{Snodgrass}, H.~B. 1983, \apj, 270, 288, \dodoi{10.1086/161121}

\bibitem[{{Svalgaard} \& {Wilcox}(1975)}]{SVAL_1975}
{Svalgaard}, L., \& {Wilcox}, J.~M. 1975, \solphys, 41, 461,
  \dodoi{10.1007/BF00154083}

\bibitem[{{Torrence} \& {Compo}(1998)}]{TOR_98}
{Torrence}, C., \& {Compo}, G.~P. 1998, Bulletin of the American Meteorological
  Society, 79, 61, \dodoi{10.1175/1520-0477(1998)079<0061:APGTWA>2.0.CO;2}

\bibitem[{{van Driel-Gesztelyi} \& {Green}(2015)}]{vanD_2015}
{van Driel-Gesztelyi}, L., \& {Green}, L.~M. 2015, Living Reviews in Solar
  Physics, 12, 1, \dodoi{10.1007/lrsp-2015-1}

\bibitem[{{Wang}(1998)}]{WANG_1998}
{Wang}, Y.~M. 1998, in Astronomical Society of the Pacific Conference Series,
  Vol. 154, Cool Stars, Stellar Systems, and the Sun, ed. R.~A. {Donahue} \&
  J.~A. {Bookbinder}, 131

\bibitem[{{Wang} \& {Sheeley}(1994)}]{WS_1994}
{Wang}, Y.~M., \& {Sheeley}, N.~R., J. 1994, \jgr, 99, 6597,
  \dodoi{10.1029/93JA02105}

\bibitem[{{Wolfram Research, Inc.}(2023)}]{MATH_2023}
{Wolfram Research, Inc.} 2023, Mathematica Online, Version 13.3.
\newblock \url{https://www.wolfram.com/mathematica}

\end{thebibliography}
\bibliographystyle{aasjournal}



\end{document}